\documentclass[a4paper,11pt]{article}
\usepackage{pos}
\usepackage{caption}
\usepackage{subcaption}

\usepackage[utf8]{inputenc}
\DeclareUnicodeCharacter{27F9}{\ensuremath{\Longrightarrow}}
\DeclareUnicodeCharacter{2209}{\ensuremath{\notin}}
\DeclareUnicodeCharacter{266D}{\ensuremath{\flat}}
\DeclareUnicodeCharacter{2208}{\ensuremath{\in}}
\DeclareUnicodeCharacter{00E9}{\'{e}}
\DeclareUnicodeCharacter{2264}{\ensuremath{\leq}}
\DeclareUnicodeCharacter{2265}{\ensuremath{\geq}}
\DeclareUnicodeCharacter{2200}{\ensuremath{\forall}}
\DeclareUnicodeCharacter{221E}{\ensuremath{\infty}}
\DeclareUnicodeCharacter{221D}{\ensuremath{\propto}}
\DeclareUnicodeCharacter{210F}{\ensuremath{\hbar}}
\DeclareUnicodeCharacter{2020}{\ensuremath{\dagger}}
\DeclareUnicodeCharacter{03B1}{\ensuremath{\alpha}}
\DeclareUnicodeCharacter{03B2}{\ensuremath{\beta}}
\DeclareUnicodeCharacter{0393}{\ensuremath{\Gamma}}
\DeclareUnicodeCharacter{03B3}{\ensuremath{\gamma}}
\DeclareUnicodeCharacter{03B8}{\ensuremath{\theta}}
\DeclareUnicodeCharacter{21D2}{\ensuremath{\Rightarrow}}
\DeclareUnicodeCharacter{0394}{\ensuremath{\Delta}}
\DeclareUnicodeCharacter{03B4}{\ensuremath{\delta}}
\DeclareUnicodeCharacter{03C1}{\ensuremath{\rho}}
\DeclareUnicodeCharacter{03BE}{\ensuremath{\xi}}
\DeclareUnicodeCharacter{03C0}{\ensuremath{\pi}}
\DeclareUnicodeCharacter{03A0}{\ensuremath{\Pi}}
\DeclareUnicodeCharacter{03BC}{\ensuremath{\mu}}
\DeclareUnicodeCharacter{03BD}{\ensuremath{\nu}}
\DeclareUnicodeCharacter{222B}{\ensuremath{\int}}
\DeclareUnicodeCharacter{2211}{\ensuremath{\Sigma}}
\DeclareUnicodeCharacter{03C3}{\ensuremath{\sigma}}
\DeclareUnicodeCharacter{03C4}{\ensuremath{\tau}}
\DeclareUnicodeCharacter{03BB}{\ensuremath{\lambda}}
\DeclareUnicodeCharacter{03B7}{\ensuremath{\eta}}
\DeclareUnicodeCharacter{03A6}{\ensuremath{\Phi}}
\DeclareUnicodeCharacter{03D5}{\ensuremath{\phi}}
\DeclareUnicodeCharacter{03A8}{\ensuremath{\Psi}}
\DeclareUnicodeCharacter{03C8}{\ensuremath{\psi}}
\DeclareUnicodeCharacter{03F5}{\ensuremath{\epsilon}}
\DeclareUnicodeCharacter{2202}{\ensuremath{\partial}}
\DeclareUnicodeCharacter{220F}{\ensuremath{\prod}}
\DeclareUnicodeCharacter{03C9}{\ensuremath{\omega}}
\DeclareUnicodeCharacter{2260}{\ensuremath{\neq}}
\DeclareUnicodeCharacter{22C5}{\ensuremath{\cdot}}
\DeclareUnicodeCharacter{27E8}{\ensuremath{\langle}}
\DeclareUnicodeCharacter{27E9}{\ensuremath{\rangle}}
\DeclareUnicodeCharacter{2192}{\ensuremath{\rightarrow}}
\DeclareUnicodeCharacter{00B2}{\ensuremath{{}^2}}
\DeclareUnicodeCharacter{221A}{\ensuremath{\sqrt{}}}
\DeclareUnicodeCharacter{223C}{\ensuremath{\sim}}
\DeclareUnicodeCharacter{226A}{\ensuremath{\ll}}
\DeclareUnicodeCharacter{00D7}{\ensuremath{\times}}

\title{Constructing Compact Ans\"{a}tze for Scattering Amplitudes}

\author*[a]{Giuseppe De Laurentis}
\author[b]{Ben Page}

\affiliation[a]{Physikalisches Institut, Albert-Ludwigs-Universität at Freiburg, \\
  Hermann-Herder.Str.~3, D-79104 Freiburg, Germany}

\affiliation[b]{Theoretical Physics Department, CERN, \\
  Geneva, Switzerland}

\emailAdd{giuseppe.de.laurentis@physik.uni-freiburg.de}
\emailAdd{ben.page@cern.ch}

\abstract{In these proceedings, we discuss the recent approach of
  Ref.~\cite{DeLaurentis:2022otd} for the construction of compact Ans\"atze for
  scattering amplitudes.
  The method builds powerful constraints on the analytic structure of the rational
  functions in amplitudes from numerical tests of their behavior close
  to singularity surfaces.
  We discuss how we systematically understand these surfaces and how the
  singular behavior of the rational function can be incorporated into an Ansatz
  using techniques from algebraic geometry.
  To perform the numerical sampling, we make use of $p$-adic numbers, a
  number-theoretical field that can be considered a cousin of finite fields. The
  $p$-adic numbers
  admit a non-trivial absolute value, as well as analytic functions such as
  the $p$-adic logarithm.
  We provide a detailed example of the approach applied to an NMHV tree
  amplitude and discuss the efficacy when applied to the two-loop
  leading-color amplitude for three-photon production at hadron colliders.
  }

\FullConference{%
  Loops and Legs in Quantum Field Theory - LL2022,\\
  25-30 April, 2022\\
  Ettal, Germany
}

\begin{document}
\maketitle

\section{Introduction}

Meeting the calculational requirements of modern hadron collider experiments is
theoretically very demanding. Due to the high precision projected at the LHC, it
is necessary to push fixed-order cross-section computations to NNLO QCD
precision for a variety of multi-scale observables, see e.g.~\cite{Huss:2022ful}.
Many such predictions require the computation of two-loop QCD scattering 
amplitudes with five external particles. In recent years, we have seen impressive
advances in computational techniques that have allowed the calculation of many
two-loop amplitudes involving five massless particles~\cite{Abreu:2018aqd,Chicherin:2018yne,Abreu:2019rpt,Chicherin:2019xeg,Badger:2018enw,Abreu:2018zmy,Abreu:2019odu,Abreu:2021oya,Badger:2019djh,Chawdhry:2019bji,Abreu:2020cwb,Chawdhry:2020for,Agarwal:2021grm,Agarwal:2021vdh,Chawdhry:2021mkw,Badger:2021imn} as well
as a number involving four massless and a single massive
particle~\cite{Abreu:2021asb, Badger:2021nhg,Badger:2021ega,Badger:2022ncb}.

In the modern approach, scattering amplitudes are computed in a divide-and-conquer
fashion, where they are broken down into a linear combination of special
functions, whose coefficients depend rationally on the external kinematics. The
special functions and rational coefficients are then computed separately.
In recent multi-scale loop calculations, this is often
considered at the level of the finite remainder. Specifically, after
ultra-violet renormalization, one can subtract away the infra-red poles to
define a finite remainder $\mathcal{H}_n^{(l)}$~\cite{Catani:1998bh, Becher:2009cu, Gardi:2009qi, Becher:2009qa}
\begin{equation}
   \mathcal{A}^{(l)}_{n,R} = \sum_{l' = 0}^{l-1} {\bf I}^{(l-l')}_\epsilon \mathcal{A}^{(l')}_{n,R} + \mathcal{H}_n^{(l)} + \mathcal{O}(\epsilon) \, .
\end{equation}
This finite remainder admits a decomposition into a basis of special
functions as
\begin{equation}
   \mathcal{H}_n^{(l)} = \sum_{i} \; {\mathcal{C}_i(\lambda, \tilde\lambda)} \; \times \; F_i(\lambda, \tilde\lambda) \, .
 \end{equation}
The special functions $F_i$ can be constructed in a way that depends
only on the kinematics of the process under study, see, e.g.~the recent
development of sets of pentagon functions~\cite{Gehrmann:2018yef,Chicherin:2020oor,Chicherin:2021dyp}.
However, the coefficients $\mathcal{C}_i$ are complicated process-dependent
rational functions. In recent years, a new paradigm has emerged which has
enabled an explosion of computations of multi-scale
amplitudes. In short, one writes an Ansatz for the $\mathcal{C}_i$, and fixes
the unknown parameters from numerical evaluations over finite fields~\cite{vonManteuffel:2014ixa,Peraro:2016wsq}.

As we look towards amplitudes which depend on an
increased number of scales, we see that the number of unknowns in the
frequently used ``common denominator'' Ansatz is growing
significantly, as one can see in Table~\ref{tab:AnsatzScalingTable}. This has
motivated us to search for a strategy for constructing improved
Ans\"atze for the rational coefficients in QCD scattering amplitudes.

\begin{center}
  \begin{tabular}{|c|cccc|} 
    \cline{2-5}
    \multicolumn{1}{c|}{} & $\mathcal{H}^{(2)}_{pp\rightarrow jjj}$ & $\mathcal{H}^{(2)}_{pp \rightarrow \gamma\gamma\gamma}$ &  $\mathcal{H}^{(2)}_{pp \rightarrow W+2j}$ &  $\mathcal{A}^{(1)}_{0\rightarrow 6g} + \mathcal{O}(\epsilon)$ \\[0.5ex]
    \hline Ref. & \cite{Abreu:2019odu} & \cite{Chawdhry:2019bji, Abreu:2020cwb} & \cite{Abreu:2021asb} & \cite{Laurentis:2019bjh} \\
    \hline Ansatz size & $\mathcal{O}(10^5)$ & $\mathcal{O}(10^5)$ & $\mathcal{O}(10^7)^{(*)}$ & $\mathcal{O}(10^9)$ \\ \hline
  \end{tabular}
  \captionsetup{justification=centering}
  \captionof{table}{Ansatz size for the numerators when written in
    common denominator form. \\ $^{(*)}$ Reduced to $\mathcal{O}(10^6)$
    after the univariate partial-fraction decomposition of
    Ref.~\cite{Badger:2021nhg}. }
  \label{tab:AnsatzScalingTable}
\end{center}

In recent years, it has been observed that if the rational prefactors
$\mathcal{C}_i$ are cast in a multivariate partial fractions
decomposition, significant cancellations occur (see
e.g.~Ref.~\cite{Heller:2021qkz}).  However, in practice, these
simplifications are achieved after first reconstructing the result in
a less compact form. In this talk, we discuss the approach of
Ref.~\cite{DeLaurentis:2022otd}, where an Ansatz inspired by the
structure of a partial fractions decomposition is constructed from
numerical evaluations in singular limits. In practice, these Ans\"atze
exhibit a reduced number of free parameters in comparison to the
corresponding common denominator form and thereby significantly reduce
the computational cost of analytic reconstruction.

\section{Ansatz Construction Approach}

To introduce our strategy, let us
consider the 6-point NMHV tree amplitude $A_{0 →
  q^+g^+g^+\bar{q}^-g^-g^-}$. We work with the assumption that we do not know
its analytical form, as if we were numerically computing a
loop-integral coefficient. The least common denominator is easily
obtained from numerical evaluations, thus we can write
\begin{equation}\label{eq:tree}
  A_{0 → q^+g^+g^+\bar{q}^-g^-g^-} = \frac{\mathcal{N}}{⟨12⟩⟨23⟩⟨34⟩[45][56][61]s_{345}} \, ,
\end{equation}
where $\mathcal{N}$ is an unknown polynomial in spinor brackets. That
is, $\mathcal{N}$ belongs to a vector space $\mathcal{M}_{d,\vec\phi}$
spanned by spinor-bracket monomials of appropriate mass dimension $d$ and
little-group weights~$\vec{\phi}$. The mass dimension (6) as well as
the little-group weights (-1, 0, 0, 1, 0, 0) of $\mathcal{N}$ are easily
obtained, thus we have
\begin{equation}
  \mathcal{N} \in \mathcal{M}_{6, [-1, 0, 0, 1, 0, 0]} \, , \quad \text{where} \quad \text{dim}(\mathcal{M}_{6, [-1, 0, 0, 1, 0, 0]}) = 143 \, .
\end{equation}
We aim to construct an Ansatz for $\mathcal{N}$, based on
the expectation that the result can be cast in some partial-fractioned
form, where the numerator has been partially cancelled against
the denominator.
For example, if we are able to show that $\mathcal{N}$ can be written
in the form
\begin{equation}
  \mathcal{N} = n_{12} \langle 12 \rangle + n_{23} \langle 23 \rangle \, ,
  \label{eq:TwoDenominatorDecomposition}
\end{equation}
then we can cancel this against factors of $\langle 12 \rangle$ and
$\langle 23 \rangle$ in the denominator. In Ref.~\cite{Laurentis:2019bjh},
it was suggested that the validity of such a decomposition can be
deduced from evaluations of the amplitude on phase-space points where
$\langle 12 \rangle$ and $\langle 23 \rangle$ are
$\mathcal{O}(\epsilon)$, for some small quantity $\epsilon$.  Note
that there are actually two separate possibilities to achieve this,
either the spinor $\lambda_2$ is nearly soft, or the spinors
$\lambda_1$, $\lambda_2$ and $\lambda_3$ are simultaneously almost
collinear.
In practice, for $A_{0 → q^+g^+g^+\bar{q}^-g^-g^-}$ one finds that
\begin{align}
  \begin{split}
    \lambda_2 &\sim \epsilon \Rightarrow \mathcal{N} \sim \epsilon^0, \\ 
    \langle 12 \rangle \sim \langle 23 \rangle \sim \langle 13 \rangle & \sim \epsilon \Rightarrow \mathcal{N} \sim \epsilon^1.
    \end{split}
\end{align}
Therefore, the tentative decomposition of
Eq.~\eqref{eq:TwoDenominatorDecomposition} does not
hold. Nevertheless, we still have found a constraint from the
vanishing of the numerator on the simultaneously collinear surface. It
can be shown that the subspace of $\mathcal{M}_{d,\vec\phi}$ that
satisfies this constraint has dimension 84, i.e.~1 evaluation fixed 59
parameters. We naturally wish to incorporate this style of constraint into an Ansatz.
To systematize this strategy, we therefore find ourselves needing to answer a
number of questions:
\begin{itemize}
  \item How do we understand the different ways in which we can set the
    denominators small?
  \item How do we numerically obtain the degree of vanishing for the
    $\mathcal{C}_i$ in two-loop amplitudes?
  \item How do we incorporate into an Ansatz the constraints provided by the
vanishing of $\mathcal{N}$?
\end{itemize}
Let us consider these problems one by one.

\subsection{Branching of singular varieties}
First, we interpret the task of numerically setting a pair of denominators
$\mathcal{D}_i$ and $\mathcal{D}_j$ to be small as finding phase-space points
which are close to the surfaces where the denominators are zero. That is, we are
interested in understanding the solutions of
\begin{equation}
  \mathcal{D}_i = \mathcal{D}_j = 0.
\end{equation}
These are surfaces defined by the zero sets of
polynomials and are known as algebraic varieties. The fact that there are
different ways of setting the denominators small corresponds to the fact that
these varieties may have multiple branches. More mathematically, varieties
associated to pairs of denominator factors may be reducible.

\begin{figure}[h]
     \centering
     \begin{subfigure}[b]{0.3\textwidth}
         \centering
         \includegraphics[width=\textwidth]{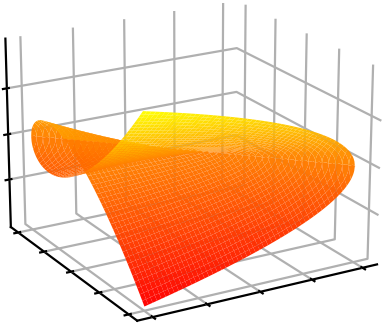}
         \caption{$V(\langle xy^2+y^3-z^2\rangle)$}
         \label{fig:V1}
     \end{subfigure}
     \hfill
     \begin{subfigure}[b]{0.3\textwidth}
         \centering
         \includegraphics[width=\textwidth]{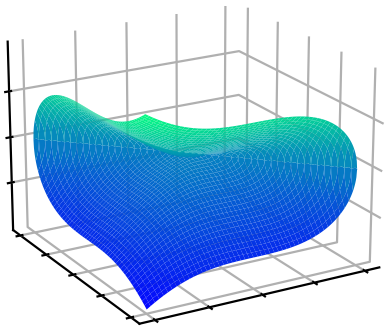}
         \caption{$V(\langle x^3+y^3-z^2 \rangle)$}
         \label{fig:V2}
     \end{subfigure}
     \hfill
     \begin{subfigure}[b]{0.3\textwidth}
         \centering
         \includegraphics[width=\textwidth]{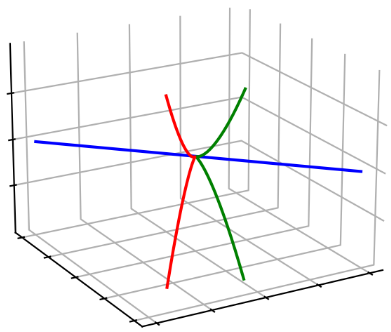}
         \caption{$V(\langle xy^2+y^3-z^2, x^3+y^3-z^2 \rangle)$}
         \label{fig:V3}
     \end{subfigure}
     \captionsetup{justification=centering}
     \caption{Two irreducible codimension-one varieties
       in Fig.~\ref{fig:V1} and Fig.~\ref{fig:V2}, \\ and the three
       branches of their reducible codimension-two intersection in
       Fig.~\ref{fig:V3}.}
     \label{figure1}
\end{figure}

A variety is said to be irreducible if it cannot be written as the
union of simpler varieties, i.e.~$U$ is irreducible if $U = U_1 \cup
U_2$ implies $U_1 = U$ or $U_2 = U$. A variety $U$ admits a
minimal decomposition in terms of irreducible varieties $U_k$,
\begin{equation}\label{eq:variety-decomposition}
  U = \bigcup_{k=1}^{n_B(U)} U_k \quad \text{s.t.} \quad U_i \not\subseteq U_j \; \forall \; i \neq j.
\end{equation}
In Fig.~\ref{figure1} we provide simple, graphical examples of
irreducible and reducible varieties in $\mathbb{R}^3$.

In higher dimensions, the reducibility of a given variety may be non-trivial to
detect from its defining polynomials. In order to systematically approach the
problem of determining the branches of a variety, we formulate the problem
algebraically.
Specifically, to the varieties $U$ and $U_k$ in the decomposition
in Eq.~\eqref{eq:variety-decomposition}, we associate the algebraic ideals, $I(U)$ and $I(U_k)$
respectively. In this language, the decomposition is translated to
\begin{equation}
  I(U) = \bigcap_{k=1}^{n_B(U)} I(U_k).
\end{equation}
The task of describing the branching of $U$ now becomes the task of finding
generators for each of the $I(U_k)$. As described in Ref.~\cite{DeLaurentis:2022otd}, this can be
achieved by applying primary decomposition techniques to the ideal $\langle
\mathcal{D}_i, \mathcal{D}_j \rangle_{R_n}$, making use of the computer algebra system
\texttt{Singular} \cite{DGPS}.

Returning to the motivating example, we see that the fact that there are two
ways to make $\langle  12 \rangle$ and $\langle 23\rangle$ simultaneously small
algebraically corresponds to the primary decomposition
\begin{equation}
  \big\langle  \langle 12 \rangle, \langle 23 \rangle  \big\rangle_{R_6} = \big\langle  \langle 12 \rangle, \langle 23 \rangle, \langle 13 \rangle \big\rangle_{R_6} \cap \langle \lambda_2^{\alpha}  \rangle_{R_6}.
\end{equation}
To be
able to numerically check if numerator decompositions of the form of
Eq.~\eqref{eq:TwoDenominatorDecomposition} apply for any pair of
denominators, it is necessary to compute the primary decompositions of
all ideals generated by such pairs. In Ref.~\cite{DeLaurentis:2022otd} we
perform this for all singularities arising from symbol letters for
two-loop five-point massless processes.

\subsection{$\boldsymbol p$-adic numbers}

Next, we discuss our approach for the sampling of the rational functions
$\mathcal{C}_i$ close to varieties where a pair of denominators are zero. To this
end we make use of the $p$-adic numbers $\mathbb{Q}_p$, which can be thought of
as a close cousin of a finite field $\mathbb{F}_p$ and inherit some of their
stability properties. A $p$-adic number $x$ can be
defined in terms of a formal series in powers of a prime number $p$,
\begin{equation}
  x = \sum_{i=\nu_p(x)}^\infty a_i p^i = a_{\nu_p(x)} p^{\nu_p(x)} + \cdots + a_{-1} p^{-1} + a_0 + a_1 p + a_2 p^2 + \cdots \, ,
\end{equation}
where $\nu_p(x)$ is called the \textit{valuation} of $x$, and
$a_{\nu_p(x)} \neq 0$. The subset of $\mathbb{Q}_p$ with $\nu_p(x)=0$
is known as the set of $p$-adic integers $\mathbb{Z}_p$.

An important feature of $\mathbb{Q}_p$ is that it admits a non-trivial absolute value, 
\begin{equation}
  |x|_p = p^{-\nu_p(x)} \quad \Longrightarrow \quad |p|_p  < |1|_p < \left|\frac{1}{p}\right|_p.
\end{equation}
In contrast to $\mathbb{F}_p$, this allows one to find numerical configurations
such that the a pair of denominators are small by building a $p$-adic
phase-space point near varieties. In practice, we 
first find a finite-finite-valued phase-space point on the 
variety exactly. Then, we \textit{lift} the $\mathbb{F}_p$ solution to an
approximate $p$-adic solution. Further details are described in Ref.~\cite{DeLaurentis:2022otd}.

With the non-trivial absolute value on $\mathbb{Q}_p$, one can
also introduce functions defined through power series. As an example,
let us discuss the $p$-adic logarithm.  This can be defined from the
Taylor expansion around $x=1$, that is
\begin{equation}
  \log_p(1+x) = \sum_{k=1}^{\infty} \frac{(-1)^{k + 1} x^k}{k} \, ,
  \label{eq:LogarithmSeries}
\end{equation}
which converges for $|x|_p < 1$, that is for $p$-adic numbers $x$ with
$\nu_p(x) \ge 1$. To continue the logarithm outside of the radius of
convergence of Eq.~\eqref{eq:LogarithmSeries}, we can make use of the
following trick. We first recall that the logarithm relates products
to sums. That is, for any $a, b$ in $\mathbb{Q}_p$
\begin{equation}
  \log_p(ab) = \log_p(a) + \log_p(b) \, .
  \label{eq:StandardLogarithmIdentity}
\end{equation}
Next, we remind ourselves of Fermat's little theorem: for an integer $w$, one
can show that
\begin{equation}
  w^{p-1} = 1 \;\, \text{mod} \;\, p \, .
\end{equation}
In $\mathbb{Z}_p$ we see that this implies that $|w^{p-1} - 1|_p < 1$.
We can therefore apply the identity of Eq.~\eqref{eq:StandardLogarithmIdentity} to rewrite
\begin{equation}
  \log_p(w) = \frac{\log_p(w^{p - 1})}{p - 1}.
  \label{eq:RadiusOfConvergenceTrick}
\end{equation}
Now the argument of the logarithm on the right hand side of
Eq.~\eqref{eq:RadiusOfConvergenceTrick} is within the radius of convergence.
We observe that computing the $p$-adic logarithm to low precision can be
implemented efficiently by computing the exponentiation in
Eq.~\eqref{eq:RadiusOfConvergenceTrick} by the approach of repeated squaring.
Given that, for large $p$, linear algebra modulo $p^k$ is very stable, we expect
applications to efficiently find linear relations between logarithms of products
of functions, e.g.~as often required in applications of the symbol calculus.

\subsection{Constructing Compact Ans\"atze}

Given the ability to classify all irreducible singular varieties associated to
the amplitude, and to $p$-adically evaluate the rational functions close to
these varieties, it remains to mathematically understand the constraint that this
places for us on the numerator. We will then use these constraints to construct
compact Ans\"atze for the rational functions.

First, we note that we make the assumption that if the numerator
vanishes to a given degree on a single such configuration, that this
is a property of the numerator on the entire variety, not just this
point.  This allows us to make use of a tool from algebraic geometry:
the so-called ``Zariski-Nagata'' theorem~\cite{Zariski1949,
  Nagata1962, EISENBUD1979157}. Schematically, the theorem states that
the set of polynomials which vanish to $k^{\text{th}}$ order on a
variety $U$ is given by the $k^{\mathrm{th}}$ ``symbolic power'' of the
ideal associated to the variety, $I(U)^{\langle k \rangle}$. That is,
for a point $(\eta^{(\epsilon)}, \tilde{\eta}^{(\epsilon)})$ near a
variety $U$ one has that
\begin{equation}
  \mathcal{N}(\eta^{(\epsilon)}, \tilde{\eta}^{(\epsilon)}) = \mathcal{O}(\epsilon^k) \;\; \Rightarrow \;\; \mathcal{N} \in I(U)^{\langle k \rangle}.
\end{equation}
The symbolic power of an ideal is strongly related to the standard power of an
ideal, and in Ref.~\cite{DeLaurentis:2022otd} we give a number of strategies for computing a generating
set of symbolic powers.

Given a collection of varieties $\mathcal{V} = \{U_1, \ldots, U_n\}$, we
determine $\kappa(\mathcal{N}, U_i)$, the degree of vanishing of $\mathcal{N}$ on
each of these varieties. This tells us that the numerator is a member of the
ideal
\begin{equation}\label{eq:gotic_J_definition}
  \mathcal{N} \in \mathfrak{J}, \quad \text{where} \quad \mathfrak{J} = \bigcap_{U \in \mathcal{V}} I(U)^{\langle  \kappa(\mathcal{N}, U) \rangle}.
\end{equation}
From knowledge of the mass dimension and little group weights of the numerator,
we now know that $\mathcal{N} \in \mathcal{M}_{d, \vec{\phi}} \cap
\mathfrak{J}$. This is a finite-dimensional vector space, in which every element
has the same vanishing properties as $\mathcal{N}$.
Therefore, we take a
basis of this space to be our Ansatz for the numerator $\mathcal{N}$. In
Ref.~\cite{DeLaurentis:2022otd} we constructed a basis of this vector space
using Gr\"{o}bner basis and linear algebra techniques.

\section{Examples}

In order to demonstrate the details of our approach, as well as its impact when
constructing compact Ans\"atze for two-loop amplitudes, we will now discuss two examples.
First we describe in detail an application to the color-ordered
tree amplitude $A_{0 → q^+g^+g^+\bar{q}^-g^-g^-}$ introduced in
Eq.~\eqref{eq:tree}.
Second, we will summarize the result of an analogous
procedure applied to two-loop finite reminders for the process $q\bar
q \rightarrow \gamma\gamma\gamma$.

\subsection{A six-point NMHV tree from its poles and zeros}

Returning to $A_{0 → q^+g^+g^+\bar{q}^-g^-g^-}$ of Eq.~\eqref{eq:tree}, we examine the
behavior of $\mathcal{N}$ on irreducible codimension-two varieties  where
pairs of invariants vanish. Specifically, we consider those 
corresponding to physical singularities, that is $⟨ij⟩$, $[ij]$ and $s_{ijk}$.
Up to permutations and parity, there are 8 inequivalent pairs of these invariants. In other words, there are 8 inequivalent, possibly reducible, codimension-two varieties. The
corresponding ideals are given by
\begin{equation}\label{eq:primary_decompositions}
  \begin{aligned}
    & J_1 \overset{!}{=} \big\langle ⟨12⟩, ⟨13⟩ \big\rangle_{R_6} = P_1 \cap P_2 \, , & \qquad & J_5 \overset{!}{=} \big\langle ⟨12⟩, [34] \big\rangle_{R_6} = P_6 \, , \\
    & J_2 \overset{!}{=} \big\langle ⟨12⟩, ⟨34⟩ \big\rangle_{R_6} = P_3 \, , & \qquad & J_6 \overset{!}{=} \big\langle ⟨12⟩, s_{123} \big\rangle_{R_6} = P_1 \cap P_7 \, , \\
    & J_3 \overset{!}{=} \big\langle ⟨12⟩, [12] \big\rangle_{R_6} = P_4 \, , & \qquad & J_7 \overset{!}{=} \big\langle ⟨12⟩, s_{134} \big\rangle_{R_6} = P_8 \, , \\
    & J_4 \overset{!}{=} \big\langle ⟨12⟩, [13] \big\rangle_{R_6} = P_5 \, , & \qquad & J_8 \overset{!}{=} \big\langle s_{123}, s_{124} \big\rangle_{R_6} = P_9 \, .
  \end{aligned}
\end{equation}
The right-hand sides provide primary decompositions for the $J_i$ and all ideals
denoted as $P_i$ are prime. The ones which do not directly correspond to
the $J_i$ ideals are
\begin{gather}
  P_1 = \big\langle ⟨12⟩, ⟨13⟩, ⟨23⟩ \big\rangle_{R_6} \, , \quad
  P_2 = \big\langle \lambda_1^\alpha \big\rangle_{R_6} \, , \quad
  P_7 = \big\langle ⟨12⟩, \lambda_1^\alpha[13]+\lambda_2^\alpha[23] \big\rangle_{R_6} \, .
\end{gather}

We now sample the numerator of the NMHV amplitude on phase space points near to the
set of codimension-two varieties. 
In practice, we observe that on 28 of these varieties the numerator vanishes. On 2 of these 28 it 
vanishes quadratically, while on the remaining ones it vanishes
linearly.
By application of Eq.~\eqref{eq:gotic_J_definition}, we have $\mathcal{N} \in
\mathfrak{J}$, where $\mathfrak{J}$ is
\begin{gather} 
  \mathfrak{J} = \Big [ P_1 \!\cap\! \overline{P}_1^{⟨2⟩}\!\!\cap\! P_1(216543)  \!\cap\! \overline{P}_1(216543) \!\cap\! P_1(432165) \!\cap\! \overline{P}_2 \!\cap\! P_3 \!\cap\! P_5(126345) \!\cap\! P_6(432165) \nonumber \\
  \phantom{\Big[} \!\cap\! P_7 \!\cap\! \overline{P}_7 \!\cap\! P_7(321654) \!\cap\! P_7(543216) \!\cap\! P_8(231654) \Big] \cap \Big[123456\rightarrow\overline{456123}\Big] \, , 
  \label{eq:ideal_membership_full}
\end{gather}
where we have made use of the anti-symmetry of the amplitude
($\overline{456123}$) to write the intersection of 28 ideals in terms
of two related sets of 14. The arguments of the ideals denote
permutations of the external legs in the one-line notation
$(\sigma(1)\dots\sigma(6))$, and an overline denotes the swap
$\lambda_{i,\alpha} \leftrightarrow \bar\lambda_{i,\dot\alpha}$. The
only second symbolic power that is needed is of the prime ideal $P_1$
(or permutations thereof). In this case, it can be shown that
$P_1^{⟨2⟩}=P_1^2$, i.e.~the second symbolic power is easily computed
as it coincides with the second (standard) power.

It is useful to write $\mathfrak{J}$ in terms of ideals generated by two
polynomials to see the connection to partial fractions. Specifically,
we can write $\mathfrak{J}$ as\footnote{We note that the argument of
  each square bracket is not in one-to-one correspondence with that
  of Eq. (21).}
\begin{align} \label{eq:ideal_membership_maximal_codim}
  \mathfrak{J} = \; & \Big[  \big\langle [12], [13], [23] \big\rangle_{R_6}^{2} \cap \big\langle ⟨12⟩, ⟨34⟩ \big\rangle_{R_6} \cap \big\langle ⟨12⟩, [16] \big\rangle_{R_6} \cap \big\langle ⟨34⟩, [12] \big\rangle_{R_6} \cap \big\langle [12], [13] \big\rangle_{R_6} \nonumber \\
  & \phantom{\Big[} \cap \big\langle [15], [16] \big\rangle_{R_6} \cap \big \langle ⟨23⟩, s_{345} \big \rangle_{R_6} \cap \big\langle ⟨12⟩, s_{123} \big\rangle_{R_6} \cap \big\langle [12], s_{123} \big\rangle_{R_6} \cap \big\langle [12], s_{345} \big\rangle_{R_6} \nonumber \\
  & \phantom{\Big[} \cap \big\langle ⟨4|2+3|1], s_{123} \big\rangle_{R_6} \cap \big\langle ⟨6|1+2|3], s_{345} \big\rangle_{R_6} \Big] \cap \Big[123456\rightarrow\overline{456123}\Big] \, ,
\end{align}
where we have also used the primary decomposition
\begin{equation}\label{eq:one-more-prime-dec}
  \big\langle ⟨1|3+4|2], s_{134} \big\rangle_{R_6} = P_1(134256) \cap P_1(\overline{256134}) \cap P_7(\overline{341562}) \cap P_7(562341) \, .
\end{equation}
Interestingly, this
involves a spinor chain despite it not being part of the initial set
of singularities.
Some of these intersections can be further carried out compactly,
\begin{align} \label{eq:ideal_membership_regroupped}
  \mathfrak{J} = \; & \Big[ \big\langle [12], [13], [23] \big\rangle_{R_6}^{2} \cap \big\langle s_{12}, ⟨34⟩[16][12], ⟨34⟩⟨12⟩[13], ⟨34⟩[13][16], [16][23]⟨34⟩ \big\rangle_{R_6} \nonumber \\
  & \phantom{\Big[} \cap \big\langle ⟨23⟩[15], ⟨23⟩[16], [16]s_{345}, [15]s_{345} \big \rangle_{R_6} \cap \big\langle ⟨12⟩[12], [12]s_{123}, s_{123}s_{345} \big\rangle_{R_6} \nonumber \\
  & \phantom{\Big[} \cap \big\langle ⟨4|2+3|1], s_{123} \big\rangle_{R_6} \cap \big\langle ⟨6|1+2|3], s_{345} \big\rangle_{R_6} \Big] \cap \Big[123456\rightarrow\overline{456123}\Big] \, .
\end{align}
Importantly, such a representation would allow 
the approach of Ref.~\cite{DeLaurentis:2022otd} to compute $\mathcal{M}_{d,
  \vec{\phi}} \cap \mathfrak{J}$ more efficiently.

Nevertheless, in this simple example, we can compute the intersection
$\mathcal{M}_{d, \vec \phi} \cap \mathfrak{J}$ directly by explicitly
computing $\mathfrak{J}$ from one of the above Eqs.~\eqref{eq:ideal_membership_full},
\eqref{eq:ideal_membership_maximal_codim} or
\eqref{eq:ideal_membership_regroupped} up to the required mass
dimension $d=6$. This can be done with $\texttt{Singular}$ by
setting an appropriate degree bound (\texttt{degBound}). The resulting
Ansatz is completely fixed, up to an overall numerical prefactor
$c_0$,
\begin{gather}
  \mathcal{M}_{d,\vec\phi} \cap \mathfrak{J} = c_0
          \big(⟨12⟩[12]⟨45⟩[45]⟨4|2+3|1]+[16]⟨34⟩⟨6|1+2|3]s_{123}\big) \, .
\end{gather}
Interestingly, we have shown that $A^{(0)}_{0 →
  q^+g^+g^+\bar{q}^-g^-g^-}$ is in fact the unique function with the
given zeros and poles in complex phase space.
In general, when considering rational coefficients of master integrals
or of special transcendental functions, the ansatz obtained from
$\mathcal{M} \cap \mathfrak{J}$ may still contain several terms,
rather than a single one. Nevertheless, the imposed constraints are
significant.

\subsection{Application to Leading-Color Two-Loop Tri-Photon Production Amplitudes}

The strategy of Ref.~\cite{DeLaurentis:2022otd} was constructed in order to
decrease the number of samples required when computing two-loop
multi-scale scattering amplitudes. In order to show the benefits of our approach
in a two-loop five-point context, we applied it to the
finite remainders of the two-loop three-photon production amplitudes
of Ref.~\cite{Abreu:2020cwb}.
Working with the analytic form of these results, we sampled them close to all
codimension-two varieties where the amplitude exhibits a (potentially spurious)
singularity. We therefore used $317$ $p$-adic evaluations to construct the
Ans\"atze.
In Table
\ref{tab:FivePointAmplitudeImprovements}, we characterize the
improvements that our strategy provides in terms of the number of free parameters of
the Ansatz. Comparing the size of the Ansatz used in
Ref.~\cite{Abreu:2020cwb}, $\mathrm{dim}(\mathcal{M}_{\tilde{d},\vec 0})$, with
the size of Ansatz in spinor variables taking into account numerator vanishing
constraints, $\mathrm{dim}(\mathcal{M}_{d,\vec {\phi}}) \cap \mathfrak{J}$, we see an
improvement of a factor of $73$ in the worst case.

\begin{table}[!htb]
    \centering
    \begin{tabular}{|l|c|c|c|c|c|c|}
    \hline
        Remainder & \begin{tabular}{@{}c@{}} vector space: \\ Mandelstam Ansatz \end{tabular} & \begin{tabular}{@{}c@{}} dim. of \\ $\mathcal{M}_{\tilde{d},\vec 0}$ \end{tabular} & \begin{tabular}{@{}c@{}} vector space: \\ spinor Ansatz \end{tabular} & \begin{tabular}{@{}c@{}} dim. of \\ $\mathcal{M}_{d, \vec \phi}$ \end{tabular} & \begin{tabular}{@{}c@{}} dim. of \\ $\mathcal{M}_{d, \vec\phi} \cap \mathfrak{J}$ \end{tabular} & ratio \\ \hline\hline
        $R^{\{2,0\}}_{\gamma^-\gamma^+\gamma^+}$ & $\mathcal{M}_{\{50,\vec 0\}}$ & 41301 & $\mathcal{M}_{35, \{3, 0, 6, -3, -2\}}$ & 7358 & 566 & 73 \\ \hline
        $R^{\{2,N_f\}}_{\gamma^-\gamma^+\gamma^+}$ & $\mathcal{M}_{\{24,\vec 0\}}$ & 2821 & $\mathcal{M}_{15, \{-2, -2, 0, -3, -3\}}$ & 378 & 20 & 141 \\ \hline
        $R^{\{2,0\}}_{\gamma^+\gamma^+\gamma^+}$ & $\mathcal{M}_{\{32,\vec 0\}}$ & 7905 & $\mathcal{M}_{20, \{-2, -4, -2, -2, -2\}}$ & 1140 & 18 & 439 \\ \hline
        $R^{\{2,N_f\}}_{\gamma^+\gamma^+\gamma^+}$ & $\mathcal{M}_{\{18,\vec 0\}}$ & 1045 & $\mathcal{M}_{8, \{1, 3, 1, 1, 2\}}$ & 44 & 6 & 174 \\ \hline
    \end{tabular}
    \captionsetup{justification=centering}
    \captionof{table}{Characterizing data for Mandelstam, spinor and refined
      spinor Ans\"atze. Spinor helicity variables alone can be seen to reduce
    the size of the Ansatz by almost an order of magnitude. The size is reduced
    by a further order of magnitude by imposing
    the numerator vanishing constraints on codimension-two varieties. }
    \label{tab:FivePointAmplitudeImprovements}
\end{table}

\section{Conclusion and Summary}

Modern methods for multi-scale two-loop scattering-amplitude computation employ
Ans\"atze and numerical sampling to determine rational coefficients of special functions.
The major bottleneck in this approach is the number of numerical samples
required, which is dictated by the number of free parameters in the Ans\"atze.
In this talk we discussed a strategy for building compact Ans\"atze.
Specifically, we build Ans\"atze that match the behavior of the rational
coefficients near varieties where amplitudes exhibit (potentially spurious)
singularities.
It turns out that these varieties may branch, and so we discussed a set of 
computational tools that one can use to describe this branching.
We determine the behavior of the coefficients near irreducible varieties from
numerical evaluations.
To perform these evaluations we introduced the $p$-adic numbers, a cousin of
finite fields which allow one to construct phase-space points close to a given
variety.
Interestingly, the $p$-adic numbers admit a theory of calculus, allowing one to
discuss functions defined through power series such as the $p$-adic logarithm.
With the behavior of the coefficients in singular regions in hand, we then
discussed how to interpret the degree of vanishing on the numerator on such
varieties algebraically. This then allows us to construct a complete Ansatz
matching this behavior.
To discuss our approach in practice, we looked at the example of a six-point
NMHV tree amplitude and built an Ansatz that required only one sample to
determine the full analytic result.
Thereafter, we demonstrated the impact of our strategy on a proof-of-concept
two-loop example, the leading-color amplitudes for three-photon production at
hadron colliders.
We look forward to applying this approach to scattering amplitudes
involving an increasingly higher number of scales in the future.

\setlength{\bibsep}{5pt}
\bibliography{main}

\end{document}